\documentclass[preprint,12pt]{elsarticle}

\usepackage{amsmath}
\usepackage{amssymb}
\usepackage{graphicx}
\usepackage{lineno}
\usepackage{booktabs}
\biboptions{numbers,square,sort&compress}
\usepackage[section]{placeins}
\usepackage[hidelinks,hypertexnames=false]{hyperref}
\date{November 5, 2026}

\begin{document}

\begin{frontmatter}

\title{Simulation Modeling and Control of an Elastic Shaping Antagonistic Variable Stiffness Actuator}

\author[inst1]{Yuanlong Ji\texorpdfstring{\fnref{fn1}}{}}
\author[inst1]{Ruizhe Jiang\texorpdfstring{\fnref{fn1}}{}}
\author[inst1]{Dongrun Jin}
\author[inst2]{Xiangyu Xie}
\author[inst2]{Junheng Lin}
\author[inst1]{Xingbang Yang\texorpdfstring{\corref{cor1}}{}}

\fntext[fn1]{These authors contributed equally to this work and share first authorship.}
\cortext[cor1]{Corresponding author}
\ead{yangxingbang@buaa.edu.cn}

\affiliation[inst1]{
  organization={School of Biological Science and Medical Engineering, Beihang University},
  city={Beijing},
  country={China}
}

\affiliation[inst2]{
  organization={School of Mechanical Engineering and Automation, Beihang University},
  city={Beijing},
  country={China}
}

\begin{abstract}
Position--stiffness regulation of antagonistic variable stiffness actuators (VSAs) involves coupled nonlinear dynamics, parameter uncertainty, and feedback states that are not directly measured. This paper addresses this challenge through a simulation-based workflow that integrates elastic shaping, model verification, and layered nonlinear control. This study builds on a previously developed actuator comprising two crossed four-bar compliant transmission elastic units. Its optimized elastic mapping provides the structural basis, with an ADAMS torque-curve root mean square error (RMSE) of $2.57\,\mathrm{N\!\cdot\!mm}$ against the target. A control-oriented analytical model is implemented in Simulink and compared with the independently executed ADAMS model under matched inputs. The maximum joint-angle RMSE across the tested cases is $4.66\times10^{-3}\,\mathrm{rad}$. A layered controller combines extended Kalman filtering for state reconstruction, feedback linearization for nominal decoupling, and adaptive sliding-mode control for residual compensation. At the end of the nominal simulation, the joint-angle error is approximately $3.14\times10^{-5}\,\mathrm{rad}$. The stiffness error is $3.07\,\mathrm{N\!\cdot\!mm/rad}$ relative to a target of $50\,\mathrm{N\!\cdot\!mm/rad}$. Parameter-variation and ablation tests assess tracking performance and the effects of the controller modules within the evaluated conditions. The resulting workflow links structural-response modeling, state estimation, and robust control assessment, supporting repeatable comparisons beyond nominal tracking alone.
\end{abstract}

\begin{highlights}
\item A prior crossed four-bar design provides the nonlinear elastic mapping.
\item ADAMS elastic-shaping results provide a reference for analytical modeling.
\item ADAMS and Simulink models are compared under identical input conditions.
\item A layered controller combines state estimation, decoupling and compensation.
\item Parameter-variation and ablation tests assess closed-loop performance.
\end{highlights}

\begin{keyword}
Variable stiffness actuator \sep Elastic shaping \sep Multibody simulation \sep Position-stiffness regulation \sep Layered nonlinear control
\end{keyword}

\end{frontmatter}

\section{Introduction}
\label{sec:introduction}

Robotic exoskeletons for assistance and rehabilitation must provide motion support while maintaining compliant physical interaction with the wearer~\cite{ref2,ref4}. This requirement creates a trade-off between motion performance and mechanical compliance at the actuated joints~\cite{ref5}. Variable stiffness actuators (VSAs) offer a way to address this trade-off by adjusting mechanical stiffness while generating motion~\cite{ref1}. However, stiffness adjustment changes the torque--deflection relation of the elastic transmission, linking stiffness variation to the motion response. This coupling limits independent regulation of stiffness and motion, so that stiffness adjustments can perturb the joint trajectory and may lead to tracking errors, oscillations, or instability. VSA design therefore requires the mechanical properties of the elastic mechanism to be considered alongside position--stiffness control requirements~\cite{ref12}.

Mechanical stiffness can be adjusted through the elastic elements and their transmission geometry~\cite{ref13}, for example by changing spring preload~\cite{ref22} or the transmission ratio~\cite{ref26}. Antagonistic actuation uses two opposing elastic branches, drawing inspiration from the coactivation of antagonist muscles~\cite{ref7}. The balance of branch torques determines the joint equilibrium position, whereas the elastic properties of both branches determine joint stiffness. English and Russell demonstrated antagonistic stiffness adjustment using nonlinear rolamite springs, illustrating the importance of branch elasticity~\cite{ref9}. Designing an antagonistic actuator thus requires both an appropriate branch arrangement and a torque--deflection relation compatible with position--stiffness regulation.

For branches containing linear springs, geometric transmission can shape the required nonlinear elastic response. Our earlier work introduced a crossed four-bar compliant transmission elastic (CFB-CTE) unit, optimized its structural parameters, and assessed its response using ADAMS~\cite{ji2026cfbcte}. The mechanism converts linear-spring behavior into a nonlinear torque--angle relation. However, this static elastic mapping alone cannot describe motion during control, which also involves coupling among the motors, elastic branches, and output joint. Generic VSA models incorporate these dynamics into nonlinear controller design~\cite{ref14}. For CFB-CTE, control simulation therefore requires an analytical dynamic model that incorporates the geometrically shaped elasticity and preserves the principal responses of the multibody mechanism model.

Control evaluation must also extend beyond nominal parameters and ideal state feedback to account for parameter mismatch and limited measurements. Feedback linearization has enabled motion--stiffness decoupling~\cite{ref15}, and simultaneous regulation has been demonstrated for bidirectional antagonistic actuators~\cite{ref38}. Nonlinear disturbance observers~\cite{ref17} and adaptive neural-network compensation~\cite{ref16} have further improved robustness to model uncertainty. Robust compensation, however, does not by itself provide the feedback states required by the controller. Encoders measure angles, but direct differentiation for velocity estimation is sensitive to noise~\cite{ref35}, while stiffness estimation depends on the transmission model~\cite{ref36}. Extended Kalman filtering combines model predictions with measurements to reconstruct states~\cite{ref37}. The effects of estimation errors on decoupling and compensation must nevertheless be assessed in closed-loop tests. The question for CFB-CTE is therefore whether model-based position--stiffness regulation remains effective under parameter mismatch and the measurement constraints represented in the simulation.

Addressing this question requires model-response checks to be linked with control tests under non-ideal conditions. Tracking results obtained within a single analytical model do not independently establish whether that model adequately represents the mechanism dynamics~\cite{sargent2003}. Multibody modeling provides a separate response reference based on components and constraints. Previous VSA research has used ADAMS and Simulink separately to examine mechanism responses and nonlinear dynamics~\cite{wang2025apvsa}. Virtual prototyping can reduce repeated physical testing and support design iteration without repeatedly manufacturing and assembling prototypes~\cite{rega2025vp}. Controlled simulation experiments also allow parameters and controller configurations to be varied explicitly to examine the sources of performance differences~\cite{barton2013}. Combining multibody-model comparisons, parameter-mismatch tests, and state-estimation assessments thus provides evidence beyond nominal tracking curves and supports repeatable comparison and refinement of control designs.

This study examines model-response consistency, robustness to parameter mismatch, and state feedback under limited measurement availability for the CFB-CTE antagonistic VSA. Independently executed ADAMS and Simulink models are compared, followed by closed-loop tests using the analytical model. The controller combines an extended Kalman filter (EKF), feedback linearization (FL), and radial basis function (RBF)-based adaptive sliding-mode control (SMC). These layers perform state reconstruction, nominal decoupling, and uncertainty compensation, respectively. The main contributions are:
\begin{enumerate}
  \item A control-oriented model connects the CFB-CTE geometric elastic mapping, actuator dynamics, and position--stiffness states. It incorporates the torque--angle characteristics established in the earlier structural study, providing a common basis for response comparison, internal-state analysis, and controller evaluation.
  \item Independently implemented ADAMS and Simulink models are compared through joint-angle responses under matched inputs. The maximum root mean square error (RMSE) across the tested cases is $4.66\times10^{-3}\,\mathrm{rad}$. This comparison quantifies differences between the analytical and multibody responses and provides a model-level basis for interpreting subsequent closed-loop simulations.
  \item A layered observation--decoupling--compensation controller is evaluated through nominal, parameter-variation, and ablation tests to examine state reconstruction, model mismatch, and controller configuration. At the end of the nominal simulation, the joint-angle error is $3.14\times10^{-5}\,\mathrm{rad}$. The stiffness error is $3.07\,\mathrm{N\!\cdot\!mm/rad}$ relative to the target of $50\,\mathrm{N\!\cdot\!mm/rad}$. In the tested velocity channels, EKF reduces estimation RMSE by over 97\% relative to direct numerical differentiation.
\end{enumerate}

Section~\ref{sec:modeling} presents the mechanism and dynamic model, and Section~\ref{sec:control} describes the controller design. Section~\ref{sec:simulation} reports the simulation results, followed by discussion and conclusions in Sections~\ref{sec:discussion} and~\ref{sec:conclusion}.

\section{System design and modeling}
\label{sec:modeling}

This section summarizes the previously developed CFB-CTE architecture and its elastic mapping, then establishes the control-oriented dynamic model. The nonlinear torque--angle relation connects the structural parameters to position--stiffness behavior. The earlier ADAMS elastic-shaping results provide the structural reference for this formulation. The resulting state-space representation relates the two motor inputs to joint position and equivalent stiffness.

\subsection{Antagonistic VSA structure}

\begin{figure}[!htbp]
  \centering
  \includegraphics[width=0.98\textwidth]{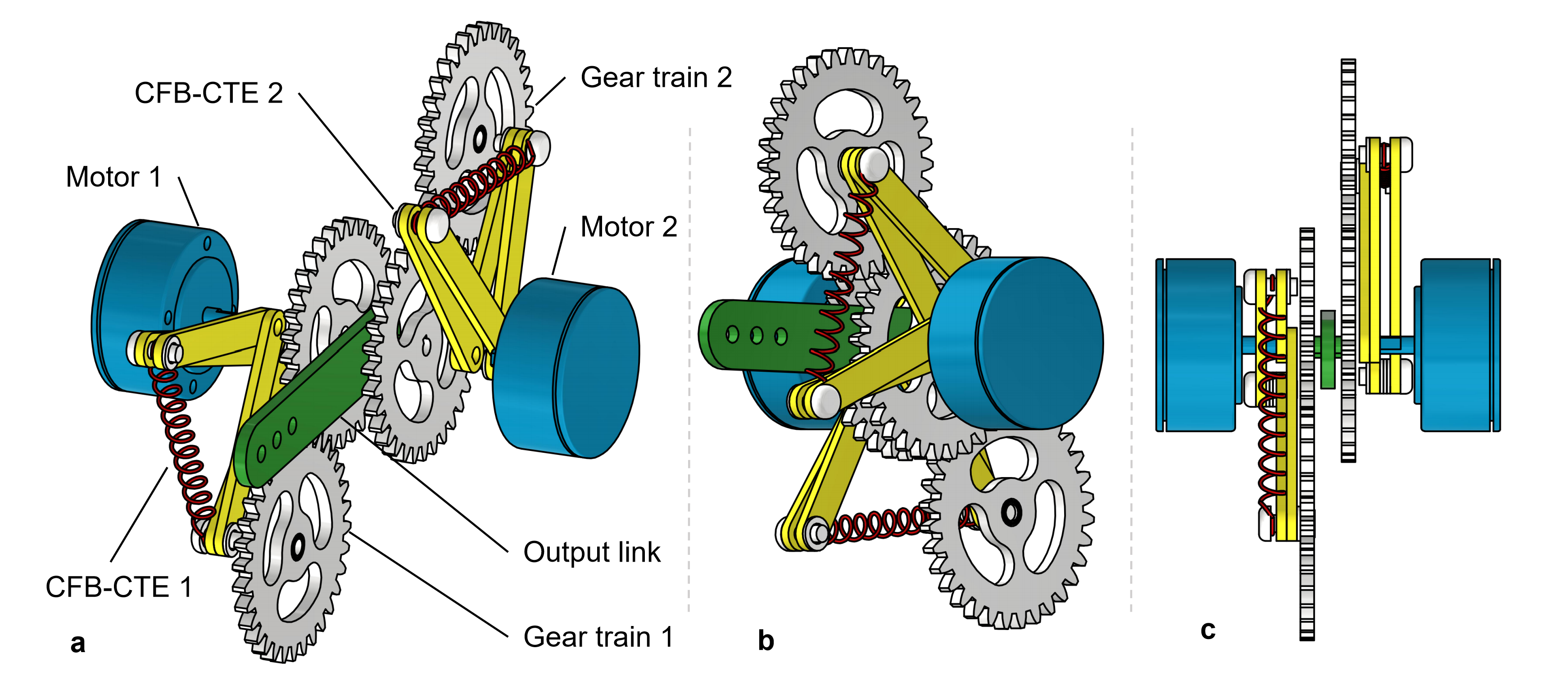}
  \caption{Structural configuration of the CFB-CTE antagonistic VSA. (a) Annotated assembly view showing two drive motors, two gear trains, two CFB-CTE units and the output link. (b) Oblique view of the antagonistic arrangement. (c) Side view showing the compact layout of the motors, gear trains and elastic transmission units.}
  \label{fig:structure}
\end{figure}

As shown in Fig.~\ref{fig:structure}, the mechanism developed in our earlier work~\cite{ji2026cfbcte} is a bio-inspired antagonistic VSA built upon two mirrored CFB-CTE units. The overall architecture comprises two identical crossed four-bar compliant transmission elastic units arranged symmetrically about the output joint, each actuated independently by a dedicated drive motor. The four-bar mechanism maps motor input angular displacement into concurrent variations in spring length and effective moment arm. The interplay of elastic force and variable moment arm gives rise to the characteristic nonlinear torque–angle relation. Thus, the CFB-CTE uses geometric nonlinearity to reshape the response of a linear spring into a nonlinear torque-angle relation.

\begin{figure}[!htbp]
  \centering
  \includegraphics[width=0.42\textwidth]{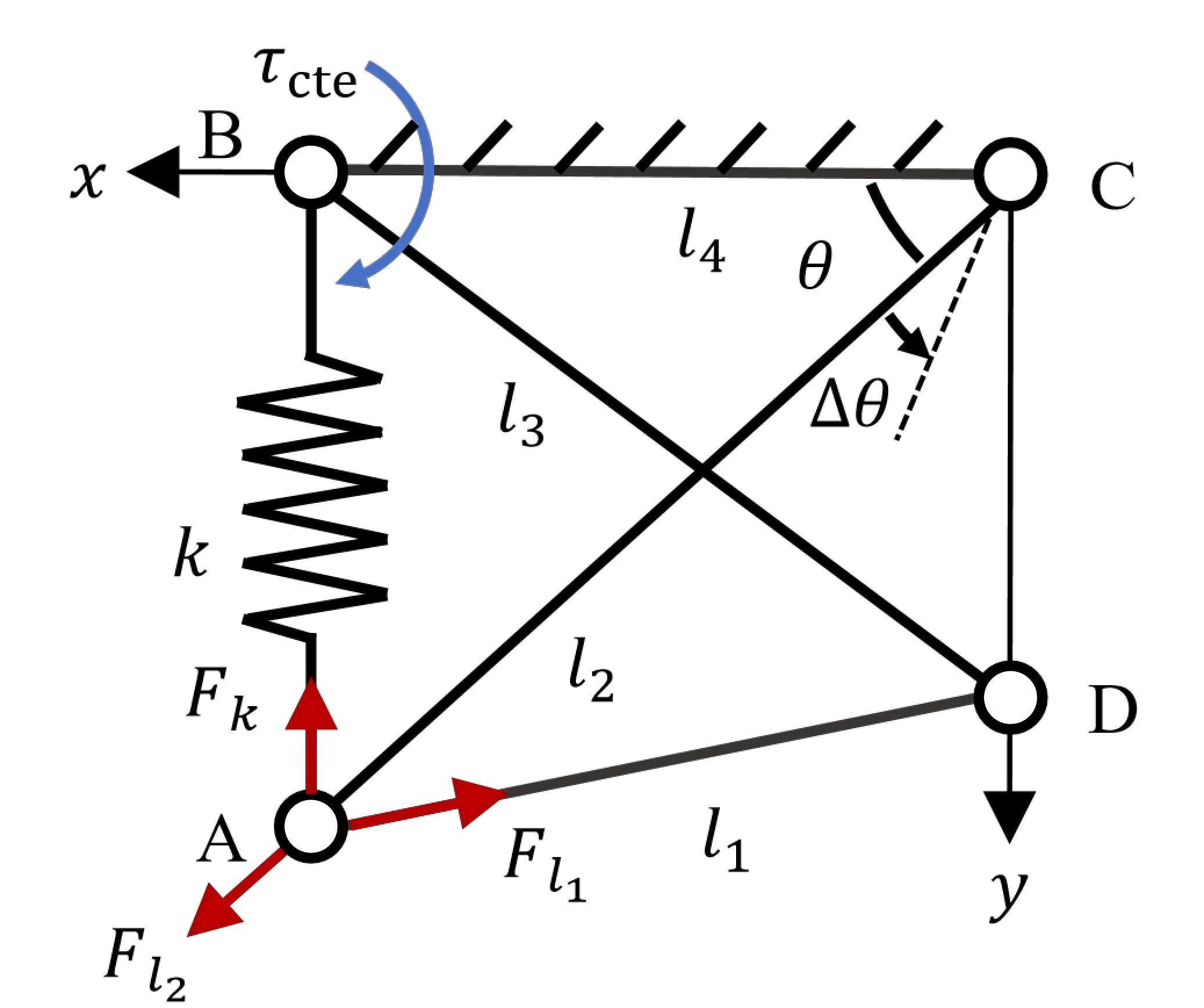}
  \caption{Parameter definition of the CFB-CTE unit. The force labels $F_k$, $F_{l1}$ and $F_{l2}$ denote the spring force and link-transmitted force components used in the geometric-elastic mapping.}
  \label{fig:structural_parameter_definition}
\end{figure}

A single CFB-CTE unit can be abstracted as the geometric-elastic mapping shown in Fig.~\ref{fig:structural_parameter_definition}. In this mapping, points $\mathrm{A}$--$\mathrm{D}$ denote the four-bar joints, $l_1$--$l_4$ denote the corresponding link lengths, $\theta$ is the input angle, and $\Delta\theta$ is the angular deflection measured from the initial installation angle. The output torque about joint $\mathrm{B}$ is denoted by $\tau_{\mathrm{cte}}$. The crossed four-bar geometry determines the spring extension and equivalent moment arm, and the output torque can be expressed as
\begin{equation}
  \tau_{\mathrm{cte}}(\theta)=F_k(\theta)r(\theta), \qquad
  F_k(\theta)=k\left[l_{\mathrm{s}}(\theta)-l_0\right].
  \label{eq:cte_torque}
\end{equation}
where $l_{\mathrm{s}}(\theta)$ is the spring length, $r(\theta)$ is the equivalent moment arm, $F_k(\theta)$ is the spring force, $k$ is the spring stiffness and $l_0$ is the spring free length. The link-force labels $F_{l1}$ and $F_{l2}$ in Fig.~\ref{fig:structural_parameter_definition} denote forces transmitted through the adjacent links. For the control-oriented representation, the equivalent elastic angular deflection is defined by $\delta=\Delta\theta=\theta-\theta_0$, where $\theta_0$ is the initial installation angle. Because $l_{\mathrm{s}}(\theta)$ and $r(\theta)$ vary nonlinearly with the input angle, the unit output in Eq.~\eqref{eq:cte_torque} can exhibit progressive nonlinear elasticity even when the spring itself is linear. The target torque-deflection relation is then approximated as
\begin{equation}
  \tau_{\mathrm{cte}}(\delta)\approx b_1\delta+a_2\delta^2, \qquad
  K_{\mathrm{cte}}(\delta)=\frac{\partial \tau_{\mathrm{cte}}}{\partial \delta}=b_1+2a_2\delta ,
  \label{eq:cte_stiffness}
\end{equation}
where $K_{\mathrm{cte}}$ is the equivalent rotational stiffness of the elastic unit. Equation~\eqref{eq:cte_stiffness} preserves the analytical form required for control modeling and reflects the stiffness-shaping role of the CFB-CTE geometry.

Two CFB-CTE units act antagonistically on the same output joint and form the left and right elastic branches. The difference between the two branch torques drives joint motion, whereas the deformation-dependent stiffness of both branches determines the equivalent joint stiffness.

This configuration determines the form of the subsequent modeling and control problem. First, the two inputs affect both position and stiffness through geometric mapping and elastic elements, resulting in coupled nonlinearity. Second, stiffness is not directly measured but is determined by elastic deformation and mechanism geometry, and thus must be explicitly modeled and estimated. Third, elasticity and transmission effects make parameter variations, friction and unmodeled dynamics directly affect the accuracy of closed-loop decoupling.

\subsection{CFB-CTE parameter optimization and ADAMS validation}

The structural optimization procedure was developed in our earlier work~\cite{ji2026cfbcte} and is summarized here to specify the mechanical basis of the control-oriented model. To generate the desired nonlinear torque-angle relation over the target working interval, the structural parameter design is formulated as a constrained target-curve fitting problem. The optimized variables are the key geometric and spring parameters listed in Table~\ref{tab:structural_parameters}, mainly including $l_1$--$l_4$, $\theta_0$, $k$ and the admissible deflection interval $\Delta\theta$. These parameters determine $l_{\mathrm{s}}(\theta)$, $r(\theta)$ and $\tau_{\mathrm{cte}}(\theta)$.

The optimization procedure is organized as follows:
\begin{enumerate}
  \item The target torque-angle curve is specified over $\theta=68^{\circ}$--$85^{\circ}$, and boundary constraints are imposed on link lengths, joint distances, spring extension, transmission angle and structural envelope.
  \item For each candidate parameter set, $l_{\mathrm{s}}(\theta)$ and $r(\theta)$ are computed to obtain the analytical torque curve through Eq.~\eqref{eq:cte_torque}.
  \item The objective function in Eq.~\eqref{eq:optimization} combines target-curve fitting error with constraints on spring preload feasibility, equivalent inertia and structural mass.
  \item The selected parameter set is imported into ADAMS to examine whether the analytical fitting curve preserves the target elastic shaping behavior in a multibody environment.
\end{enumerate}
The weighted objective is written as
\begin{equation}
  J =
  w_{\mathrm{T}}\mathrm{RMSE}(\tau_{\mathrm{cte}},\tau_{\mathrm{d}})
  +w_{\mathrm{J}}\frac{I_{\mathrm{eq}}}{I_{\mathrm{ref}}}
  +w_{\mathrm{M}}\frac{m}{m_{\mathrm{ref}}}
  +P_{\mathrm{c}} ,
  \label{eq:optimization}
\end{equation}
where $\tau_{\mathrm{d}}$ is the target torque curve, $I_{\mathrm{eq}}$ and $m$ are the equivalent inertia and structural mass, $P_{\mathrm{c}}$ summarizes constraint penalties, and $w_{\mathrm{T}}$, $w_{\mathrm{J}}$ and $w_{\mathrm{M}}$ are weighting coefficients. The preload-related constants constrain the initial spring length; material-density parameters determine the link and gear masses; and the derived quantities $r_{\mathrm{p}}$, $I_{\mathrm{g}}$ and $M_{\mathrm{tot}}$ appear in the equivalent inertia and mass terms used in the objective function and subsequent dynamics.

\begin{figure}[!htbp]
  \centering
  \includegraphics[width=0.92\textwidth]{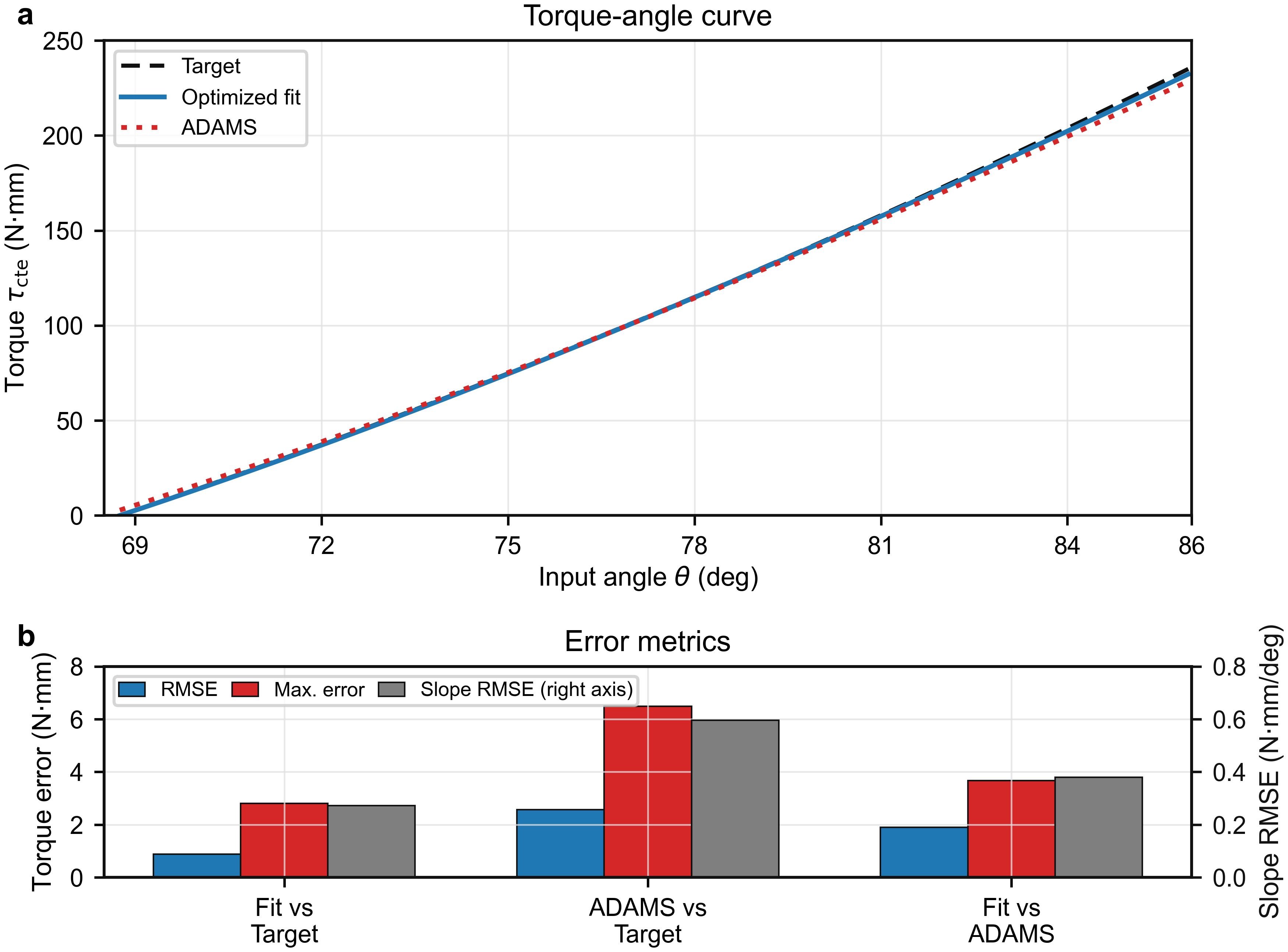}
  \caption{Nonlinear elastic shaping and ADAMS validation of the CFB-CTE unit. (a) Comparison among the target torque-angle curve, optimized analytical curve and ADAMS multibody curve in the design working interval. (b) Error statistics for analytical fitting and multibody validation. The optimized CFB-CTE parameters reproduce the intended nonlinear elastic shaping effect in the multibody model.}
  \label{fig:cte_validation}
\end{figure}

The ADAMS-based elastic-shaping validation reported in our earlier work~\cite{ji2026cfbcte} is summarized here as the structural basis for the subsequent dynamic model. As shown in Fig.~\ref{fig:cte_validation}, the target curve, optimized analytical curve and ADAMS curve maintain similar trends in the target working interval. The RMSE between the optimized analytical curve and the target curve is $0.883\,\mathrm{N\!\cdot\!mm}$, and the RMSE between the ADAMS curve and the target curve is $2.57\,\mathrm{N\!\cdot\!mm}$. These results indicate that the CFB-CTE uses crossed-four-bar geometric mapping to reshape the response of a linear spring into the target nonlinear output.

The validation supports two observations about the structure. First, the CFB-CTE geometric parameters form the intended nonlinear elastic shaping effect in the target working interval. Second, the structural response in the ADAMS multibody model remains consistent with the target curve, allowing the elastic mapping to serve as the basis for system modeling. After unit-level curve validation, the modeling focus shifts to the joint-level system arising from the antagonistic connection—specifically, to establishing the state-space representation that relates the two motor inputs and elastic pre-deformation to the resulting joint position and equivalent stiffness. The geometric and stiffness parameters of the CFB-CTE unit—including $l_1$--$l_4$, $k$, $\theta_0$ and $\Delta\theta$—jointly determine the spring extension, equivalent moment arm, and nonlinear torque–angle mapping over the target working interval. These parameters are summarized in Table~\ref{tab:structural_parameters}, with their geometric definitions illustrated in Fig.~\ref{fig:structural_parameter_definition} to support reproducibility of the simulations.

\begin{table}[!htbp]
\centering
\caption{Key parameters used for CFB-CTE structural optimization and validation. The geometric symbols are defined in Fig.~\ref{fig:structural_parameter_definition}.}
\label{tab:structural_parameters}
\footnotesize
\setlength{\tabcolsep}{4pt}
\begin{tabular}{llll}
\toprule
Symbol & Description & Unit & Value \\
\midrule
$a_2$ & Quadratic coefficient of target torque curve & N$\cdot$mm/rad$^2$ & 510.66 \\
$b_1$ & Linear coefficient of target torque curve & N$\cdot$mm/rad & 630.25 \\
$\Delta\theta$ & Equivalent angular-deflection interval & rad & 0--0.3 \\
$w_{\mathrm{T}}$ & Torque-curve fitting weight & -- & 0.9 \\
$w_{\mathrm{J}}$ & Inertia-related weight & -- & 0 \\
$w_{\mathrm{M}}$ & Mass-related weight & -- & 0.1 \\
$l_1$ & Link 1 length & mm & 63.487 \\
$l_2$ & Link 2 length & mm & 70.386 \\
$l_3$ & Link 3 length & mm & 70.000 \\
$l_4$ & Base link length & mm & 63.876 \\
$\theta_0$ & Initial installation angle & rad & 1.2000 \\
$k$ & Equivalent extension-spring stiffness & N/mm & 0.0300 \\
$r_{\mathrm{p}}$ & Gear pitch radius & mm & 31.938 \\
$I_{\mathrm{g}}$ & Polar moment of inertia of one gear & kg$\cdot$mm$^2$ & 15.87 \\
$M_{\mathrm{tot}}$ & Total mass of optimized mechanism & kg & 0.0502 \\
\bottomrule
\end{tabular}
\end{table}

\subsection{Dynamic modeling and problem formulation}
\label{sec:dynamic_model}

\begin{figure}[!htbp]
  \centering
  \includegraphics[width=0.58\textwidth]{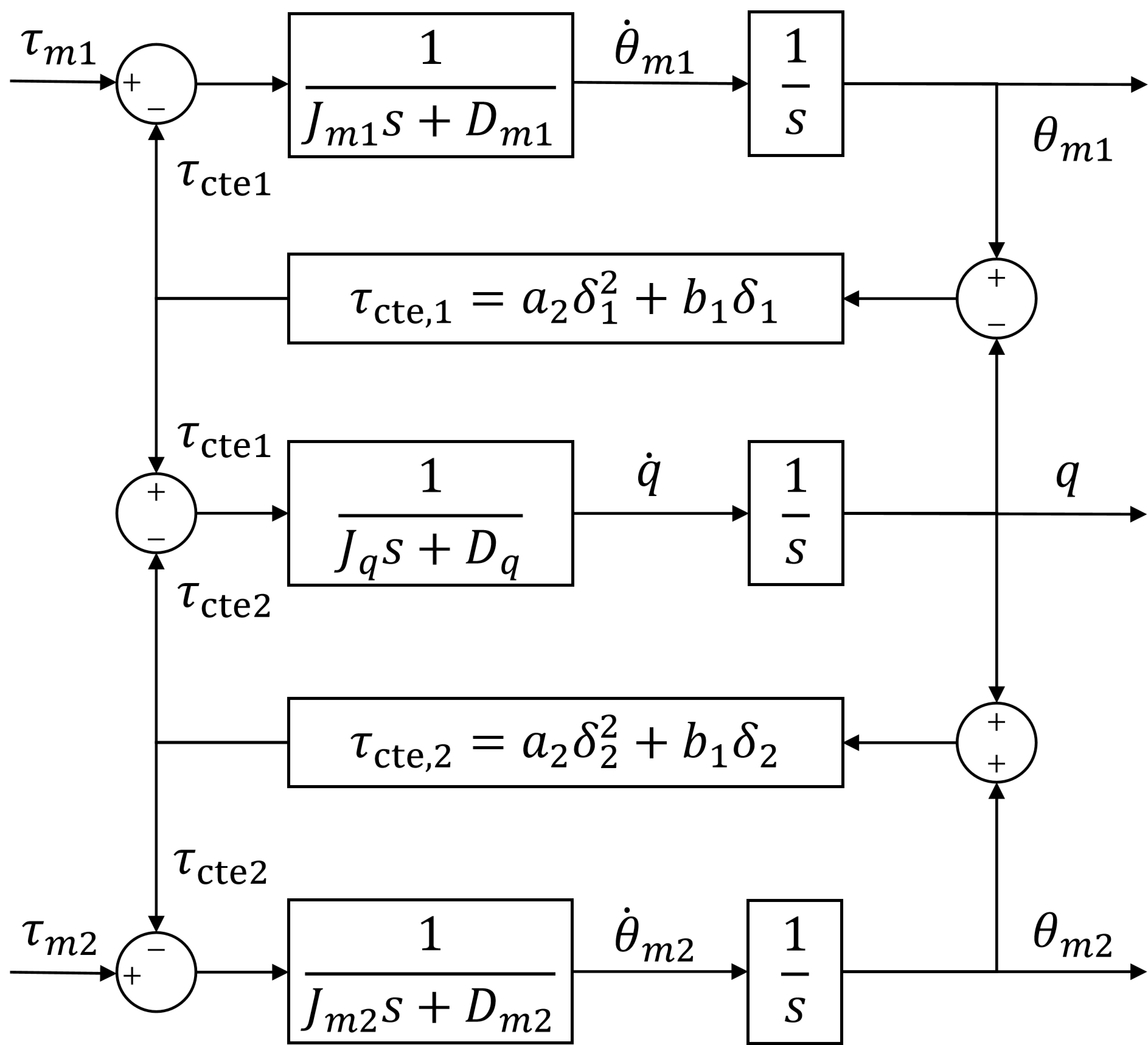}
  \caption{Analytical block representation of the CFB-CTE antagonistic VSA. The two motor torques $\tau_{\mathrm{m}1}$ and $\tau_{\mathrm{m}2}$ drive the motor-side coordinates $\theta_{\mathrm{m}1}$ and $\theta_{\mathrm{m}2}$; the joint coordinate is $q$; and the CFB-CTE elastic torques $\tau_{\mathrm{cte},1}$ and $\tau_{\mathrm{cte},2}$ are generated from the relative branch deflections $\delta_1$ and $\delta_2$ and act on both motor-side and joint-side dynamics.}
  \label{fig:dynamic_model_block}
\end{figure}

After the CFB-CTE elastic mapping is verified, the system model is built around the two-sided antagonistic connection. The parameter-optimized and ADAMS-validated CFB-CTE unit is represented as a nonlinear elastic element that links motor-side deformation with joint torque and equivalent stiffness. The motor-torque vector $\boldsymbol{\tau}_{\mathrm{m}}=[\tau_{\mathrm{m}1},\tau_{\mathrm{m}2}]^{\mathrm T}$ is transformed into differential and common-mode inputs, $\boldsymbol{\tau}_{qK}=[\tau_q,\tau_K]^{\mathrm T}$, through $\mathbf{H}_2$:
\begin{equation}
  \boldsymbol{\tau}_{qK}=\mathbf{H}_2\boldsymbol{\tau}_{\mathrm m},\qquad
  \mathbf{H}_2=\frac{1}{2}
  \begin{bmatrix}
  1 & -1\\
  1 & 1
  \end{bmatrix}.
  \label{eq:input_transform}
\end{equation}
Here, $\tau_q$ and $\tau_K$ denote the differential and common-mode torque inputs, respectively. The motor torques and angles follow the local positive directions in Fig.~\ref{fig:dynamic_model_block}, so that positive rotation of either motor increases its branch deflection at a fixed joint position. Accordingly, $\tau_q=(\tau_{\mathrm{m}1}-\tau_{\mathrm{m}2})/2$ is associated with joint-position regulation, whereas $\tau_K=(\tau_{\mathrm{m}1}+\tau_{\mathrm{m}2})/2$ is associated with elastic preload and stiffness regulation. This input transformation does not itself eliminate the nonlinear coupling between position and stiffness.

Within the dynamic and control model, $\theta$ denotes the motor differential coordinate. Together with joint position $q$, equivalent stiffness $K$ and their time derivatives, it forms the state $\boldsymbol{x}=[q,\dot q,\theta,\dot\theta,K,\dot K]^{\mathrm T}$. The continuous-time dynamics are represented in nonlinear affine form:
\begin{equation}
  \dot{\boldsymbol{x}}=
  \boldsymbol{f}_{\mathrm c}(\boldsymbol{x})+
  \mathbf{g}(\boldsymbol{x})\boldsymbol{\tau}_{qK}+
  \boldsymbol{d},
  \label{eq:affine}
\end{equation}
where $\boldsymbol{f}_{\mathrm c}$ includes inertial, CFB-CTE elastic shaping and geometric mapping terms, $\mathbf{g}$ is the input-gain matrix, and $\boldsymbol{d}$ represents parameter uncertainty, friction and unmodeled dynamics. The CFB-CTE elastic relation is embedded in $\boldsymbol{f}_{\mathrm c}$ and $\mathbf{g}$ in Eq.~\eqref{eq:affine}.

The desired position and stiffness trajectories are denoted by $q_{\mathrm d}(t)$ and $K_{\mathrm d}(t)$. The control objective is to maintain position--stiffness tracking under parameter variations. Signed output deviations used for performance evaluation are $q-q_{\mathrm d}$ and $K-K_{\mathrm d}$; the controller's error signal is defined by the reference-minus-feedback summation in Fig.~\ref{fig:control_framework}.

The corresponding analytical block representation is shown in Fig.~\ref{fig:dynamic_model_block}. The two motor-side coordinates $\theta_{\mathrm{m}1}$ and $\theta_{\mathrm{m}2}$ and the joint coordinate $q$ are connected through two CFB-CTE elastic branches. With these local motor coordinates, positive joint rotation reduces the first branch deflection and increases the second. The branch deflections are therefore defined as
\begin{equation}
  \delta_1=\theta_{\mathrm{m}1}-q,\qquad
  \delta_2=\theta_{\mathrm{m}2}+q,
  \label{eq:branch_deflection}
\end{equation}
and the corresponding branch torques are written as
\begin{equation}
  \tau_{\mathrm{cte},1}=a_2\delta_1^2+b_1\delta_1,\qquad
  \tau_{\mathrm{cte},2}=a_2\delta_2^2+b_1\delta_2 .
  \label{eq:branch_torque}
\end{equation}
Thus, the single-unit torque-deflection relation in Eq.~\eqref{eq:cte_stiffness} is embedded into the antagonistic joint-level model through the branch deflections in Eq.~\eqref{eq:branch_deflection} and the elastic torques $\tau_{\mathrm{cte},1}$ and $\tau_{\mathrm{cte},2}$. As shown in Fig.~\ref{fig:dynamic_model_block}, the net elastic torque acting on the joint is $\tau_{\mathrm{cte},1}-\tau_{\mathrm{cte},2}$, while each motor receives the opposing elastic reaction $-\tau_{\mathrm{cte},i}$ ($i=1,2$). $J_{\mathrm{m}1}$, $J_{\mathrm{m}2}$ and $J_q$ denote motor-side and joint-side inertias, while $D_{\mathrm{m}1}$, $D_{\mathrm{m}2}$ and $D_q$ denote the corresponding viscous damping terms. In nominal (baseline) simulations, the two motor-side inertia and damping blocks share the baseline motor parameters unless parameter scaling is applied.

\section{Controller design}
\label{sec:control}

\subsection{Control framework}
\label{sec:control_framework}

\begin{figure}[!htbp]
  \centering
  \includegraphics[width=\textwidth]{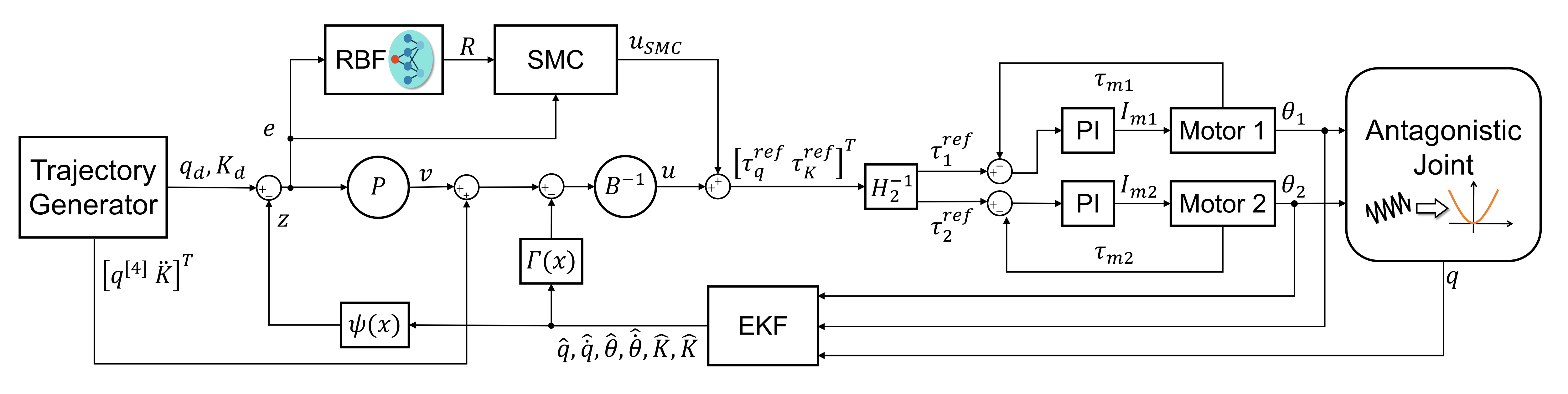}
  \caption{EKF-FL-RBF-SMC layered control framework. State estimation supplies feedback for error shaping and nominal nonlinear compensation. The nominal command and adaptive sliding-mode compensation are combined before allocation to the two motor loops.}
  \label{fig:control_framework}
\end{figure}

The CFB-CTE antagonistic VSA is a dual-input dual-output nonlinear affine system. The controller must address state unavailability, channel coupling and model residuals. The design follows a layered order: first obtain consistent state estimates, then eliminate the main nominal coupling, and finally compensate residual uncertainty. The resulting state-observation, nominal-decoupling and residual-compensation framework is shown in Fig.~\ref{fig:control_framework}.

The layered design follows the CFB-CTE structural mapping. In Fig.~\ref{fig:control_framework}, the motor angles are $\theta_1\equiv\theta_{\mathrm{m}1}$ and $\theta_2\equiv\theta_{\mathrm{m}2}$, using the local coordinates of Fig.~\ref{fig:dynamic_model_block}. The EKF output is $\hat{\boldsymbol{x}}=[\hat q,\hat{\dot q},\hat\theta,\hat{\dot\theta},\hat K,\hat{\dot K}]^{\mathrm T}$. The mapping $\boldsymbol{z}=\boldsymbol{\psi}(\hat{\boldsymbol{x}})$ supplies the reconstructed feedback signal. The controller error is $\boldsymbol{e}=[q_{\mathrm d},K_{\mathrm d}]^{\mathrm T}-\boldsymbol{z}$, and the error-feedback block $P$ produces $\boldsymbol{v}$. The trajectory generator supplies the corresponding reference derivatives; its labels $q^{[4]}$ and $\ddot K$ denote $q_{\mathrm d}^{(4)}$ and $\ddot K_{\mathrm d}$ on that feedforward branch. These terms form $\boldsymbol{r}_{\mathrm{ff}}=[q_{\mathrm d}^{(4)},\ddot K_{\mathrm d}]^{\mathrm T}$. Nominal compensation uses $\boldsymbol{\Gamma}(\hat{\boldsymbol{x}})$ and the inverse decoupling matrix $\mathbf{B}^{-1}(\hat{\boldsymbol{x}})$. The blocks marked $\psi(x)$, $\Gamma(x)$ and $B^{-1}$ are evaluated using the state estimate.

The signal $\boldsymbol{u}$ is the nominal command before residual compensation. Adding $\boldsymbol{u}_{\mathrm{SMC}}$ gives the equivalent reference input $\boldsymbol{\tau}_{qK}^{\mathrm{ref}}=[\tau_q^{\mathrm{ref}},\tau_K^{\mathrm{ref}}]^{\mathrm T}$. The inverse allocation $[\tau_1^{\mathrm{ref}},\tau_2^{\mathrm{ref}}]^{\mathrm T}=\mathbf{H}_2^{-1}\boldsymbol{\tau}_{qK}^{\mathrm{ref}}$ provides the two motor-torque references. The proportional--integral (PI) blocks regulate the motor-side loops, with $I_{\mathrm{m}1}$ and $I_{\mathrm{m}2}$ denoting motor currents in amperes and $\tau_{\mathrm{m}1}$ and $\tau_{\mathrm{m}2}$ denoting the corresponding torque-feedback signals. Reference torques and applied motor torques are kept distinct in the model.

The control-oriented model uses the equivalent inertia and damping parameters summarized in Table~\ref{tab:control_parameters}. The torque coefficients $b_1$ and $a_2$ are defined in Table~\ref{tab:structural_parameters} and are used in the branch torque model of Eq.~\eqref{eq:branch_torque}. Parameter-variation cases are generated by scaling the nominal parameter set in simulation, while the equivalent differential inertia $J_{\theta}$ and stiffness-channel inertia $J_K$ define the scaling of the position and stiffness channels in the feedback-linearization layer.

\begin{table}[!htbp]
\centering
\caption{Control-oriented model parameters used in the controller and simulation settings.}
\label{tab:control_parameters}
\footnotesize
\setlength{\tabcolsep}{4pt}
\begin{tabular}{llll}
\toprule
Symbol & Description & Value & Unit \\
\midrule
$J_{\mathrm{m}}$ & Nominal motor inertia & $7.1\times10^{3}$ & kg$\cdot$mm$^2$ \\
$J_q$ & Nominal joint inertia & $1.4\times10^{2}$ & kg$\cdot$mm$^2$ \\
$D_q$ & Nominal joint damping & $2.5\times10^{-1}$ & N$\cdot$mm\,s/rad \\
$J_{\theta}$ & Equivalent angular-difference inertia & $1.42\times10^{3}$ & kg$\cdot$mm$^2$ \\
$J_K$ & Equivalent stiffness-channel inertia & $6.95$ & kg$\cdot$mm$^2$ \\
\bottomrule
\end{tabular}
\end{table}

\subsection{EKF observer}
\label{sec:ekf_observer}

The role of EKF is to connect the structural model with measurable signals. Encoders provide joint- and motor-related angles, but the control law also requires velocity, differential variables and equivalent stiffness. Direct differentiation introduces noise into nonlinear compensation terms, while model prediction alone may deviate under parameter mismatch. The measurement vector used for the observer is defined as
\begin{equation}
  \boldsymbol{z}_k=
  \begin{bmatrix}
  q_{\mathrm{meas}} & \theta_{\mathrm{meas}} & K_{\mathrm{meas}} & \dot q_{\mathrm{meas}} & \dot\theta_{\mathrm{meas}}
  \end{bmatrix}^{\mathrm T},
  \label{eq:measurement}
\end{equation}
where $q_{\mathrm{meas}}$ is the measured joint angle, $\theta_{\mathrm{meas}}$ is the measured motor differential coordinate, $K_{\mathrm{meas}}$ is the stiffness pseudo-measurement computed from the CFB-CTE elastic geometry, and $\dot q_{\mathrm{meas}}$ and $\dot\theta_{\mathrm{meas}}$ are velocity-related measurement channels used for correction. The estimator measurement vector $\boldsymbol{z}_k$ collects these correction channels, whereas $\boldsymbol{z}=\boldsymbol{\psi}(\hat{\boldsymbol{x}})$ is the reconstructed feedback signal supplied to the controller.

\begin{figure}[!htbp]
  \centering
  \includegraphics[width=0.82\textwidth]{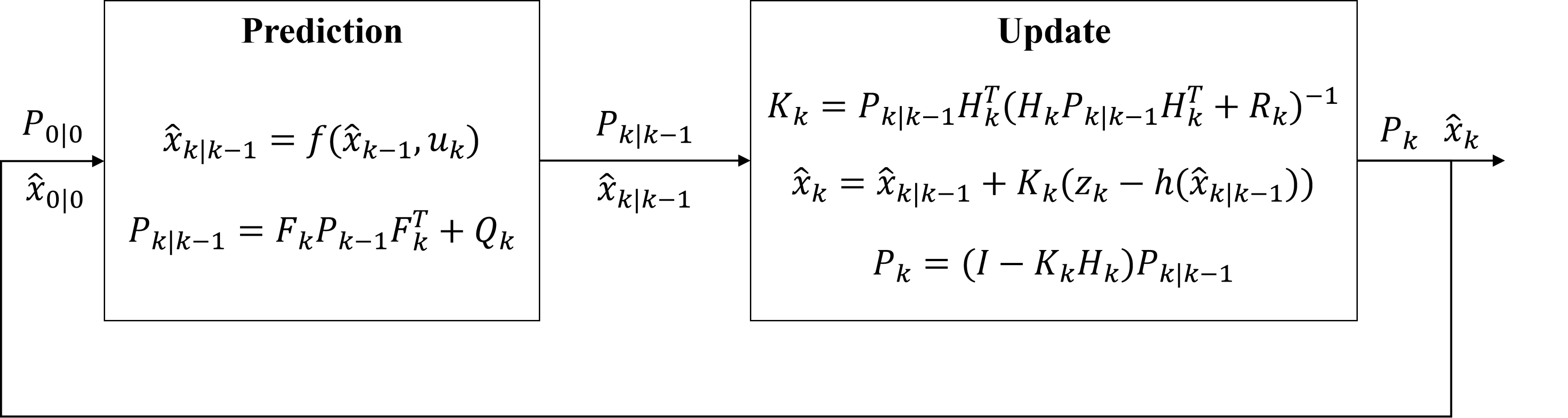}
  \caption{Prediction--update workflow of the EKF observer. The observer recursively combines nonlinear model prediction with measurement-residual correction to provide the state estimate and covariance used by the subsequent control layers.}
  \label{fig:ekf_prediction_update}
\end{figure}

The discrete nonlinear state and observation models associated with Eq.~\eqref{eq:measurement} are
\begin{equation}
  \boldsymbol{x}_k=
  \boldsymbol{f}(\boldsymbol{x}_{k-1},\boldsymbol{u}_k)+\boldsymbol{w}_k,\qquad
  \boldsymbol{z}_k=
  \boldsymbol{h}(\boldsymbol{x}_k)+\boldsymbol{v}_k,
  \label{eq:ekf_model}
\end{equation}
Here, $\boldsymbol{f}$ is the discrete state-transition function obtained from the continuous dynamics in Eq.~\eqref{eq:affine}, and $\boldsymbol{h}$ is the observation function. The index on $\boldsymbol{u}_k$ identifies the transition from $t_{k-1}$ to $t_k$: it is the applied equivalent plant input over that interval. The process and measurement noises are $\boldsymbol{w}_k$ and $\boldsymbol{v}_k$, with covariance matrices $\mathbf{Q}_k$ and $\mathbf{R}_k$, respectively. Figure~\ref{fig:ekf_prediction_update} shows the prediction--update recursion~\cite{ref37}. The posterior quantities are abbreviated as $\hat{\boldsymbol{x}}_k\equiv\hat{\boldsymbol{x}}_{k|k}$ and $\mathbf{P}_k\equiv\mathbf{P}_{k|k}$; prediction gives $\hat{\boldsymbol{x}}_{k|k-1}$ and $\mathbf{P}_{k|k-1}$. The Jacobian $\mathbf{F}_k$ is evaluated from $\boldsymbol{f}$ at $(\hat{\boldsymbol{x}}_{k-1},\boldsymbol{u}_k)$, and $\mathbf{H}_k$ is evaluated from $\boldsymbol{h}$ at $\hat{\boldsymbol{x}}_{k|k-1}$. The measurement residual is $\boldsymbol{z}_k-\boldsymbol{h}(\hat{\boldsymbol{x}}_{k|k-1})$, and $\mathbf{K}_k$ is the Kalman-gain matrix. The recursion starts from $\hat{\boldsymbol{x}}_{0|0}$ and $\mathbf{P}_{0|0}$; $\mathbf{I}$ denotes the identity matrix. Time-indexed $\mathbf{P}_k$ and $\mathbf{K}_k$ are estimator matrices, distinct from the controller block $P$ and mechanical stiffness $K$.

The covariance settings follow the relative confidence of measurements and model prediction. Angle measurement channels are assigned smaller observation noise, stiffness pseudo-measurements retain higher uncertainty to avoid over-trusting the structural model, and velocity and stiffness-rate states use larger process noise to absorb friction, parameter scaling and unmodeled dynamics. The EKF output $\hat{\boldsymbol{x}}$ is used by both feedback linearization and RBF-SMC.

\subsection{Feedback linearization}

Feedback linearization provides a model-based route to motion--stiffness decoupling \cite{ref15}. Here, it handles the nominal coupling in the CFB-CTE model. The two motor torques do not act independently on $q$ and $K$; they first pass through the differential/common-mode transformation in Eq.~\eqref{eq:input_transform} and then affect the joint through the CFB-CTE elastic mapping. Let the controlled-output vector be $\boldsymbol{o}=[q,K]^{\mathrm{T}}$. After differentiating the outputs according to their relative degrees, the system can be written as
\begin{equation}
  \boldsymbol{o}^{(r)}=
  \boldsymbol{\Gamma}(\boldsymbol{x})+
  \mathbf{B}(\boldsymbol{x})\boldsymbol{\tau}_{qK}+
  \boldsymbol{\Delta}(\boldsymbol{x},t),
  \label{eq:fl_system}
\end{equation}
where $\boldsymbol{\Gamma}$ collects input-independent nominal nonlinear terms, $\mathbf{B}$ is the decoupling matrix, and $\boldsymbol{\Delta}$ represents residuals caused by parameter variations and unmodeled dynamics. The derivative vector is $\boldsymbol{o}^{(r)}=[q^{(4)},\ddot K]^{\mathrm T}$, with the corresponding reference derivatives supplied by $\boldsymbol{r}_{\mathrm{ff}}$ in Fig.~\ref{fig:control_framework}. The matrix $\mathbf{B}$ contains the influence of CFB-CTE elastic shaping and antagonistic input mapping on the control channels.

The feedback-linearizing input is designed for the nominal part of Eq.~\eqref{eq:fl_system}. The residual term is addressed by the compensator in Section~\ref{sec:rbf_smc}:
\begin{equation}
  \boldsymbol{u}=
  \mathbf{B}^{-1}(\hat{\boldsymbol{x}})
  \left[
  \boldsymbol{v}+\boldsymbol{r}_{\mathrm{ff}}-
  \boldsymbol{\Gamma}(\hat{\boldsymbol{x}})
  \right],
  \label{eq:fl_input}
\end{equation}
where $\hat{\boldsymbol{x}}$ comes from EKF and $\boldsymbol{v}$ is the error-feedback contribution generated by $P$. Its sum with $\boldsymbol{r}_{\mathrm{ff}}$ is the complete virtual input to nominal decoupling. If the nominal model is sufficiently accurate and $\mathbf{B}$ is nonsingular in the working interval, Eq.~\eqref{eq:fl_input} cancels the nominal nonlinear terms for the equivalent input channel. The residual compensation is added downstream of this inversion, as shown in Fig.~\ref{fig:control_framework}.

Feedback linearization depends on model accuracy and the availability of suitable state estimates. Agreement with ADAMS provides a model-level response check, but does not eliminate parameter mismatch or estimation error. The feedback-linearization layer therefore targets nominal coupling, while the adaptive compensator addresses residual uncertainty within the tested conditions.

\subsection{RBF adaptive sliding-mode compensation}
\label{sec:rbf_smc}

RBF-SMC compensates residuals after feedback linearization rather than replacing the structural model and nominal decoupling. The error vector $\boldsymbol{e}=[e_q,e_K]^{\mathrm T}$ follows the reference-minus-feedback convention defined in Section~\ref{sec:control_framework}. The sliding variable is
\begin{equation}
  \boldsymbol{s}=
  \boldsymbol{\Lambda}\boldsymbol{e}
  +\int_0^t\boldsymbol{\Pi}\boldsymbol{e}(\xi)\,\mathrm{d}\xi .
  \label{eq:sliding}
\end{equation}
The positive gain matrices $\boldsymbol{\Lambda}$ and $\boldsymbol{\Pi}$ weight instantaneous and accumulated errors, respectively. The SMC block uses the sliding variable in Eq.~\eqref{eq:sliding}, while the RBF network receives the channel-wise error-state input $\boldsymbol{\xi}_i$.

Using Gaussian radial basis functions, as employed for neural approximation in VSA control \cite{ref16}, the present network estimates an adaptive gain for each channel:
\begin{equation}
  R_i=
  \boldsymbol{W}_i^{\mathrm T}\boldsymbol{h}_i(\boldsymbol{\xi}_i),\qquad
  h_{ij}=
  \exp\left(
  -\frac{\|\boldsymbol{\xi}_i-\boldsymbol{c}_{ij}\|^2}{2b_{ij}^2}
  \right),
  \label{eq:rbf}
\end{equation}
where $\boldsymbol{\xi}_i$ is the error-state input for channel $i$, $\boldsymbol{c}_{ij}$ and $b_{ij}$ are the RBF center and width, and $\boldsymbol{W}_i$ is the online-updated weight vector. The channel gains form $\boldsymbol{R}=[R_q,R_K]^{\mathrm T}$, labeled $R$ on the connection from RBF to SMC in Fig.~\ref{fig:control_framework}. A leakage term is used to limit weight drift. Adding the residual compensation to the nominal command gives the equivalent reference input
\begin{equation}
  \boldsymbol{\tau}_{qK}^{\mathrm{ref}}=
  \begin{bmatrix}\tau_q^{\mathrm{ref}}\\\tau_K^{\mathrm{ref}}\end{bmatrix}
  =\boldsymbol{u}+\boldsymbol{u}_{\mathrm{SMC}},
  \label{eq:total_input}
\end{equation}
where $\boldsymbol{u}_{\mathrm{SMC}}$ is generated from the sliding variable, boundary-layer function and RBF adaptive gain. Equation~\eqref{eq:total_input} corresponds to the summation node before $\mathbf{H}_2^{-1}$ in Fig.~\ref{fig:control_framework}. The resulting $\tau_1^{\mathrm{ref}}$ and $\tau_2^{\mathrm{ref}}$ are the references for the two motor loops. A Lyapunov function composed of sliding variables and weight-estimation errors gives $\dot V\leq0$ under bounded uncertainty and a sufficient reaching condition, ensuring bounded closed-loop errors.

\section{Simulation verification}
\label{sec:simulation}

\subsection{Simulation setup}

The simulation study separately examines model-response consistency and closed-loop performance. Independently executed ADAMS and Simulink models are first compared through joint-angle responses under matched inputs. The controller is then evaluated using the Simulink analytical model through nominal tracking, state estimation, parameter variation, and ablation tests. Model comparison covers single-input, opposite-direction dual-input, and same-direction dual-input cases, using RMSE to quantify response differences. The nominal step test uses $q_{\mathrm{d}}=0.1\,\mathrm{rad}$ and a stiffness reference, with baseline parameters listed in Tables~\ref{tab:structural_parameters} and~\ref{tab:control_parameters}. The nominal, $+5\%$, and $-5\%$ parameter cases assess sensitivity to the specified model variations. State-estimation and ablation tests examine the effects of feedback reconstruction and controller configuration. These tests evaluate the implemented simulation conditions, rather than the full range of hardware uncertainties.

\subsection{Model validation}

\begin{figure}[!htbp]
  \centering
  \includegraphics[width=0.92\textwidth]{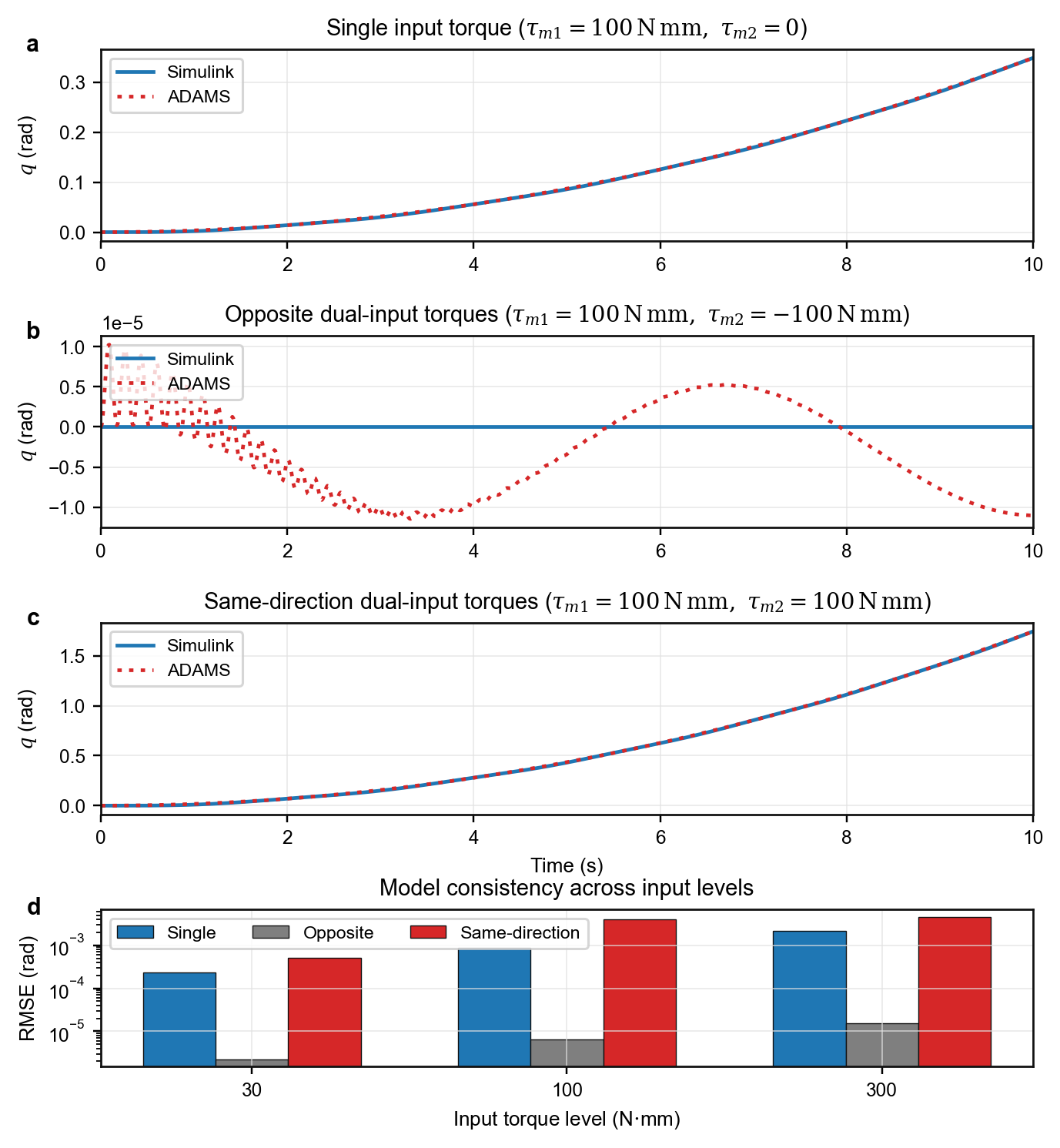}
  \caption{Consistency verification between the ADAMS multibody model and the Simulink analytical model. (a) Joint-angle response under single-input torque. (b) Joint-angle response under opposite-direction dual-input torque. (c) Joint-angle response under same-direction dual-input torque. (d) RMSE summary under three input amplitudes. The three cases cover single-branch drive, antagonistic drive and cooperative drive, and examine whether the analytical model preserves the main response features of the multibody model under different torque-distribution modes.}
  \label{fig:model_validation}
\end{figure}

To assess response consistency, the Simulink analytical model is compared with the ADAMS multibody model under three input conditions, as shown in Fig.~\ref{fig:model_validation}. In the single-input case, one motor input is fixed and step torques of different amplitudes are applied to the other motor. The opposite-direction dual-input case applies equal-magnitude step torques with opposite signs. The same-direction dual-input case applies equal-magnitude step torques with the same sign.

The three cases cover single-motor input and two dual-motor torque distributions. The analytical and ADAMS curves show consistent main trends. The RMSE ranges from $2.38\times10^{-4}$ to $2.15\times10^{-3}\,\mathrm{rad}$ in the single-input case, from $2.21\times10^{-6}$ to $1.53\times10^{-5}\,\mathrm{rad}$ in the opposite-direction case, and from $5.06\times10^{-4}$ to $4.66\times10^{-3}\,\mathrm{rad}$ in the same-direction case. These results quantify the agreement between the analytical and multibody responses over the tested inputs, providing a reference for subsequent analytical-model control simulations.

\subsection{Nominal step response}

\begin{figure}[!htbp]
  \centering
  \includegraphics[width=0.92\textwidth]{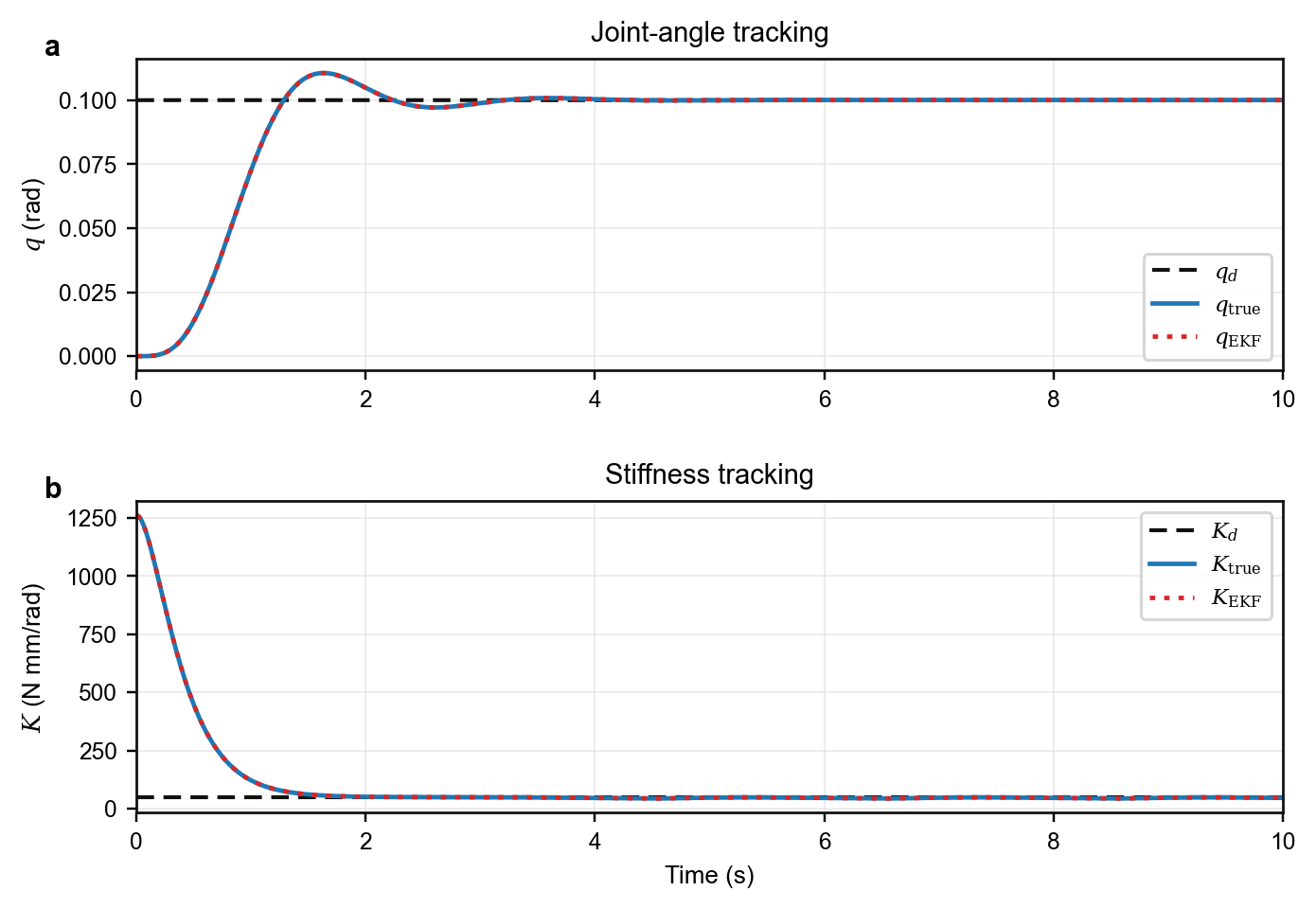}
  \caption{Position-stiffness tracking under nominal conditions. (a) Desired joint angle, simulated true response and EKF estimate. (b) Desired stiffness, simulated true response and EKF estimate. The two channels evaluate whether the closed-loop system maintains joint-position tracking and stiffness regulation under nominal parameters.}
  \label{fig:nominal}
\end{figure}

The nominal test evaluates joint-position and stiffness regulation under step references, as shown in Fig.~\ref{fig:nominal}. The simulation duration is $10\,\mathrm{s}$ and the sampling frequency is $100\,\mathrm{kHz}$. The joint-angle reference is $0.1\,\mathrm{rad}$, and the stiffness reference switches from an initial high-stiffness state to a lower target stiffness interval.

The position channel enters the target-angle neighborhood and maintains stable output, while the stiffness channel enters the target compliant interval. The final joint angle is $0.100031\,\mathrm{rad}$, corresponding to a deviation of about $3.14\times10^{-5}\,\mathrm{rad}$ from the $0.1\,\mathrm{rad}$ reference. The final stiffness is $46.93\,\mathrm{N\!\cdot\!mm/rad}$, corresponding to a deviation of about $3.07\,\mathrm{N\!\cdot\!mm/rad}$ from the target value of $50\,\mathrm{N\!\cdot\!mm/rad}$. During the last $1\,\mathrm{s}$, the mean absolute joint-angle deviation is about $3.82\times10^{-5}\,\mathrm{rad}$ and the mean absolute stiffness deviation is about $1.54\,\mathrm{N\!\cdot\!mm/rad}$. The true and EKF-estimated values remain close in both channels, with EKF RMSE values over the same final one-second interval of about $3.31\times10^{-5}\,\mathrm{rad}$ and $1.60\times10^{-2}\,\mathrm{N\!\cdot\!mm/rad}$, respectively.

\subsection{State estimation}
\label{sec:state_estimation}

\begin{figure}[!htbp]
  \centering
  \includegraphics[width=0.92\textwidth]{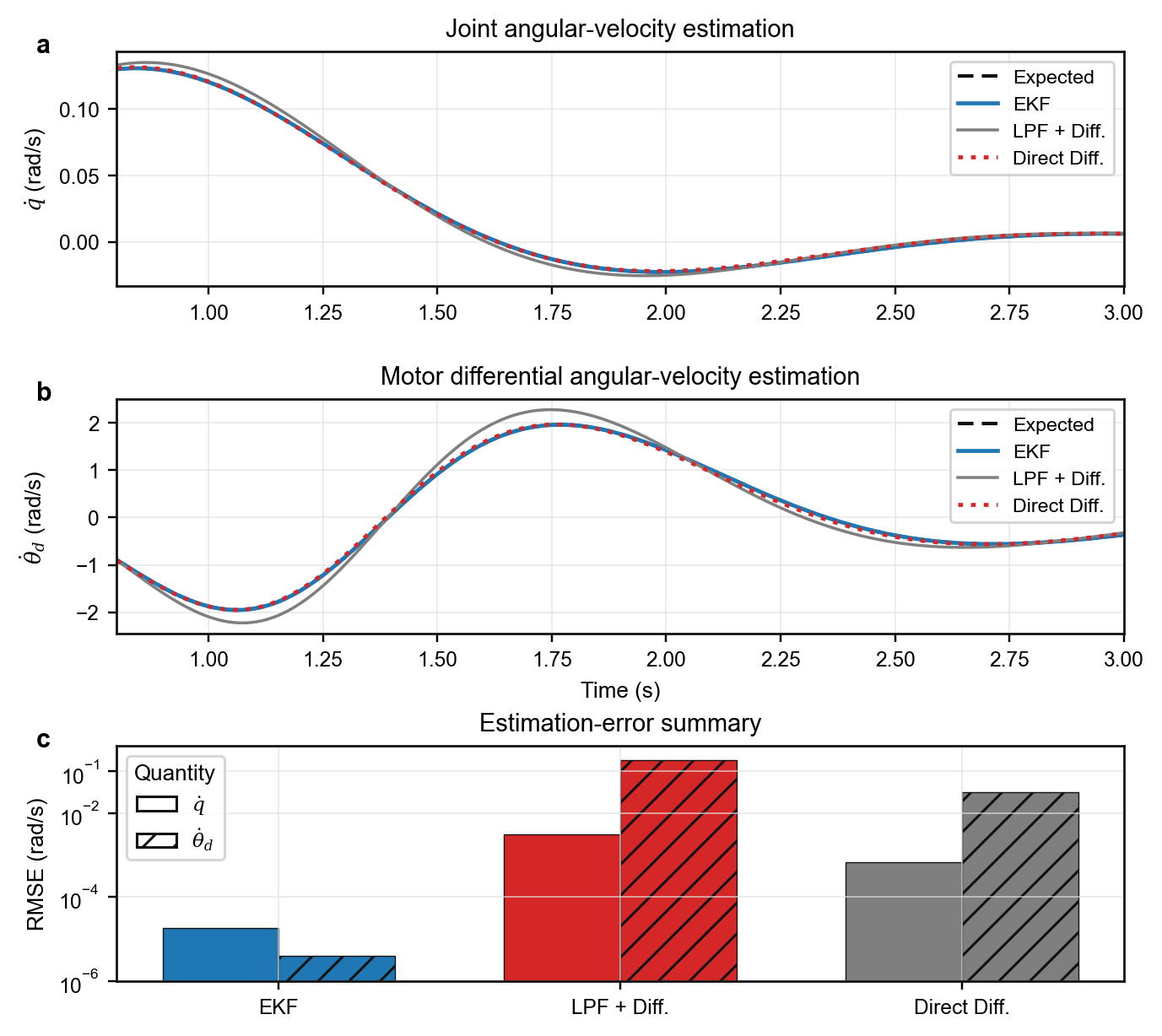}
  \caption{Comparison between EKF and differentiation-based velocity-estimation methods. (a) Joint angular-velocity estimation. (b) Motor differential angular-velocity estimation. (c) RMSEs of the three methods. Compared with LPF + Diff. and Direct Diff., EKF provides velocity estimates that remain more consistent with the reference velocity trends.}
  \label{fig:ekf}
\end{figure}

EKF is compared with low-pass filter plus differentiation (LPF + Diff.) and direct differentiation (Direct Diff.) for velocity estimation, as shown in Fig.~\ref{fig:ekf}. The compared variables are joint angular velocity $\dot q$ and motor differential angular velocity $\dot\theta$. The differential-coordinate subscript in the label $\dot\theta_{\mathrm d}$ in Fig.~\ref{fig:ekf} refers to this same $\dot\theta$.

Direct differentiation is more sensitive to angle-measurement noise, while low-pass filtering reduces high-frequency fluctuation at the cost of amplitude consistency in fast-varying segments. EKF combines dynamic prediction and measurement correction, and yields lower RMSE values for both variables. The EKF RMSE for joint angular velocity is $1.83\times10^{-5}\,\mathrm{rad/s}$, whereas direct differentiation and LPF + Diff. give $6.67\times10^{-4}\,\mathrm{rad/s}$ and $3.15\times10^{-3}\,\mathrm{rad/s}$, respectively. For motor differential angular velocity, the EKF RMSE is $3.99\times10^{-6}\,\mathrm{rad/s}$, compared with $3.12\times10^{-2}\,\mathrm{rad/s}$ and $1.83\times10^{-1}\,\mathrm{rad/s}$ for direct differentiation and LPF + Diff. These results support the use of EKF velocity estimates in the subsequent feedback-linearization layer for the tested conditions.

\subsection{Parameter variation}

\begin{figure}[!htbp]
  \centering
  \includegraphics[width=0.92\textwidth]{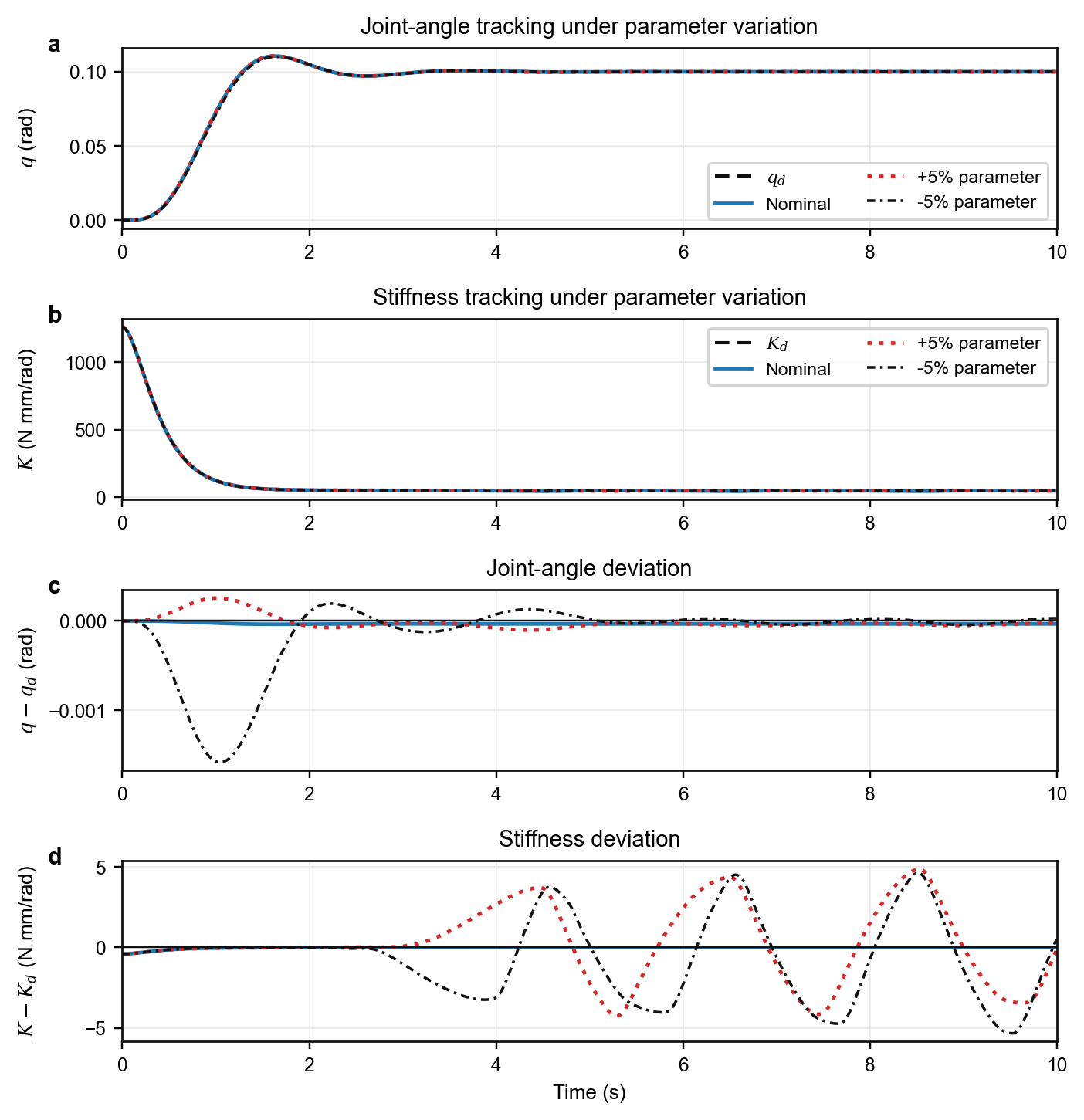}
  \caption{Position-stiffness tracking and error distribution under parameter variations. (a) Joint-angle trajectories under nominal, $+5\%$ and $-5\%$ parameter settings. (b) Stiffness trajectories under the same settings. (c) Joint-angle deviation from the reference. (d) Stiffness deviation from the reference. The results show that position tracking is maintained under parameter variation, while stiffness deviation remains bounded.}
  \label{fig:robustness}
\end{figure}

To evaluate robustness to parameter mismatch, non-nominal parameter cases are compared with the nominal case, as shown in Fig.~\ref{fig:robustness}. The parameter settings include nominal, uniformly $+5\%$ and uniformly $-5\%$ cases.

The position channel shows relatively low sensitivity to uniform parameter variations. Over the full simulation interval, the position-error RMSE is $7.88\times10^{-5}\,\mathrm{rad}$ for the $+5\%$ case and $4.18\times10^{-4}\,\mathrm{rad}$ for the $-5\%$ case. During the last $1\,\mathrm{s}$, both non-nominal cases keep the position deviation below $4.95\times10^{-5}\,\mathrm{rad}$. The stiffness channel is more affected by parameter variation, which is consistent with its dependence on elastic deformation, geometric mapping and spring parameters. During the last $1\,\mathrm{s}$, the stiffness deviations in the $+5\%$ and $-5\%$ cases remain within $3.46\,\mathrm{N\!\cdot\!mm/rad}$ and $5.34\,\mathrm{N\!\cdot\!mm/rad}$, respectively, and the final deviations are $0.152\,\mathrm{N\!\cdot\!mm/rad}$ and $0.510\,\mathrm{N\!\cdot\!mm/rad}$. Thus, parameter variation mainly changes the stiffness-error amplitude, while the closed-loop system keeps the position error small and drives stiffness into the target working interval.

\subsection{Ablation study}

\begin{figure}[!htbp]
  \centering
  \includegraphics[width=0.92\textwidth]{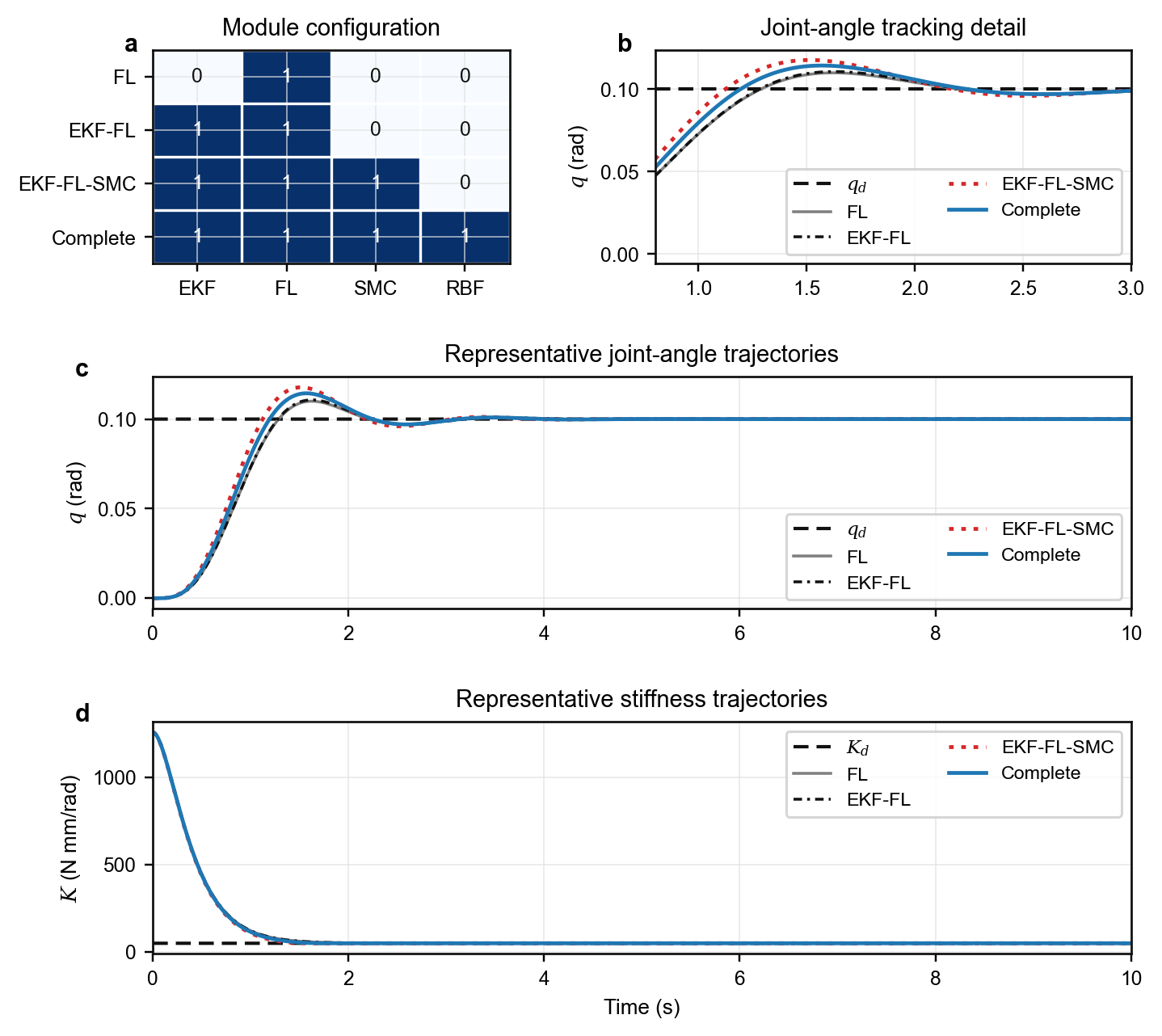}
  \caption{Ablation study of the control framework. (a) Module composition of four controller variants. (b) Local enlarged view of joint-angle tracking. (c) Representative joint-angle trajectories. (d) Representative stiffness trajectories. The comparison separates the roles of state observation, nominal decoupling, sliding-mode compensation and RBF adaptive compensation.}
  \label{fig:ablation}
\end{figure}

To analyze module contributions, four controller configurations are compared: FL only, FL + EKF, FL + EKF + SMC, and the proposed complete controller, as shown in Fig.~\ref{fig:ablation}.

The FL-only configuration mainly relies on nominal model accuracy. Adding EKF improves the consistency of state input and smooths position and stiffness trajectories. The SMC term is intended to reduce residual tracking errors. The complete controller further introduces RBF adaptive gain so that the compensation intensity changes with the error state. The full-interval position and stiffness RMSE values of FL only are $2.58\times10^{-2}\,\mathrm{rad}$ and $198.54\,\mathrm{N\!\cdot\!mm/rad}$, respectively, whereas the complete controller gives $2.53\times10^{-2}\,\mathrm{rad}$ and $197.77\,\mathrm{N\!\cdot\!mm/rad}$. The complete controller gives slightly lower full-interval RMSEs in this comparison. The trajectory plots provide complementary evidence of how the configurations differ, without establishing a universal advantage for every module.

\section{Discussion}
\label{sec:discussion}

The study links the previously developed CFB-CTE elastic mapping to dynamic modeling and closed-loop evaluation. The crossed four-bar geometry converts linear-spring behavior into a nonlinear torque--angle relation, which is retained in the analytical model. Comparison with the independently implemented ADAMS model quantifies response differences under matched inputs. This provides evidence of model consistency within the tested conditions, rather than direct validation against physical measurements.

The separation of multibody-response checks and analytical-model control tests enables repeatable comparisons of parameters and controller configurations. ADAMS represents the mechanism and its constraints, whereas the Simulink model supports state estimation and controller evaluation. The two models are executed independently, without online software coupling. Their agreement may still reflect shared assumptions or omitted effects; it does not establish hardware fidelity or closed-loop performance on the multibody plant.

The control tests address parameter mismatch and feedback reconstruction within the implemented simulation settings. EKF supplies state estimates, FL handles nominal coupling, and RBF-SMC adjusts the residual compensation gain. The stiffness channel is more sensitive to parameter variation than the position channel in the reported cases. These observations support further evaluation of the layered design, but do not establish robustness to arbitrary disturbances or sensor imperfections. Future work will prioritize online coupling of the controller with a higher-fidelity multibody model. Hardware tests and additional cases involving external disturbances, actuator limits, and sensor effects will then examine how far the simulated behavior extends to physical operation.

\section{Conclusion}
\label{sec:conclusion}

This paper developed a simulation workflow for evaluating position--stiffness regulation of the CFB-CTE antagonistic VSA under model uncertainty and limited feedback availability. Building on the earlier elastic-shaping design, it connected analytical dynamics, multibody-response comparisons, and layered state-estimation and control tests. The independently executed ADAMS and Simulink models showed consistent principal responses under the evaluated inputs. Closed-loop simulations on the analytical model quantified tracking and state-estimation errors and assessed parameter variation and controller ablations. The results provide a repeatable basis for assessing model response, feedback reconstruction, and parameter-mismatch sensitivity, with conclusions restricted to the tested simulation conditions.

\section*{Acknowledgments}
The authors acknowledge the School of Biological Science and Medical Engineering, Beihang University; the School of Engineering Medicine, Beihang University; and the School of Mechanical Engineering and Automation, Beihang University, for providing academic and research support. The authors also acknowledge the Key Laboratory of Biomechanics and Mechanobiology (Beihang University), Ministry of Education; the Key Laboratory of Innovation and Transformation of Advanced Medical Devices, Ministry of Industry and Information Technology; and the National Medical Innovation Platform for Industry-Education Integration in Advanced Medical Devices (Interdiscipline of Medicine and Engineering), Beihang University, for supporting the related research environment.

\section*{Funding}
This research was supported by the National Natural Science Foundation of China (Grant No. 52475291), the Beijing Natural Science Foundation (Grant No. L222139), the Fundamental Research Funds for the Central Universities (Grant Nos. GW2025-71 and JK2026-MS-39), and the 2025 Doctoral Student Special Program of the Young Elite Scientists Sponsorship Program by CAST.

\section*{CRediT authorship contribution statement}
Yuanlong Ji: Conceptualization, methodology, simulation modeling, data analysis, visualization, and writing -- original draft. Ruizhe Jiang: Methodology, simulation modeling, data curation, validation, visualization, and writing -- original draft. Dongrun Jin: Model implementation, validation, formal analysis, and manuscript revision. Xiangyu Xie: Mechanical design analysis, simulation validation, and manuscript revision. Junheng Lin: Mechanical design analysis, methodology discussion, and manuscript revision. Xingbang Yang: Conceptualization, supervision, project administration, funding acquisition, and writing -- review and editing.

\section*{Declaration of competing interest}
The authors declare that they have no known competing financial interests or personal relationships that could have appeared to influence the work reported in this paper.

\section*{Data availability}
Data will be made available on request.

\section*{Declaration of generative AI and AI-assisted technologies}
The authors declare that no generative AI or AI-assisted technologies were used in the scientific design, simulations, data analysis, interpretation of results, writing, or editing of this manuscript.

\end{document}